# *In Silico* Screening of Some Naturally Occurring Bioactive Compounds Predicts Potential Inhibitors against SARS-COV-2 (COVID-19) Protease*

Ashok Kumar Mishra[1] and Satya Prakash Tewari

*Department of Physics, Dr.Shakuntala Misra National Rehabilitation University, Lucknow, Uttar Pradesh, India-226017*

[1] Corresponding Author: akmishra@dsmnru.ac.in; akmishra2k5@gmail.com (A.K. Mishra)

- This research is dedicated to the peoples who have lost the lives, been struggling for the lives, been working hard to save the lives as well as been frightened for the lives while fighting against the pandemic COVID-19 i.e. the all CORONA WARRIORS.

# Abstract

SARS-COV-2 identified as COVID-19 in Wuhan city of China in the month of December, 2019 has now been declared as pandemic by World Health Organization whose transmission chain and cure both have emerged as a tough problem for the medical fraternity. The reports pertaining to the treatment of this pandemic are still lacking. We firmly believe that Nature itself provides a simple solution for any complicated problem created in it which motivated us to carry out *In Silico* investigations on some bioactive natural compounds reportedly found in the fruits and leaves of *Anthocephalus Cadamba* which is a miraculous plant found on the earth aiming to predict the potential inhibitors against aforesaid virus. Having modeled the ground state ligand structure of the such nine natural compounds applying density functional theory at B3LYP/631+G (d, p) level we have performed their molecular docking with SARS-COV-2 protease to calculate the binding affinity as well as to screen the binding at S-protein site during ligand-protein interactions. Out of these nine studied naturally occurring compounds; Oleanic Acid has been appeared to be potential inhibitor for COVID-19 followed by Ursolic Acid, IsoVallesiachotamine,Vallesiachotamine,Cadambine,Vincosamide-N-Oxide, Isodihydroamino-cadambine, Pentyle Ester of Chlorogenic Acid and D-Myo-Inositol. Hence these bioactive natural compounds or their structural analogs may be explored as anti-COVID19 drug agent which will be possessing the peculiar feature of cost-less synthesis and less or no side effect due to their natural occurrence. The solubility and solvent-effect related to the phytochemicals may be the point of concern. The *In-vivo* investigations on these proposed natural compounds or on their structural analogs are invited for designing and developing the potential medicine/vaccine for the treatment of COVID-19 pandemic.

**Introduction**

The novel corona virus disease (COVID-19) identified in Wuhan City of China [1] spread across the earth as pandemic putting the whole world on high alert [2–5] led to 823626 total cases and 40598 deaths all over the world till 01 April 2020 [6]. Available evidence indicates that this virus is transmitted through the respiratory droplets (such as coughing) and by fomites that can propagate through air at distances of 1 meter [7-9] and no evidence is available about its airborne transmission in the current study [10] which establishes the principle of maintaining the distance of more than 1 meter termed as 'social distancing'/'physical distancing' along with hand-hygiene for the prevention of COVID-19. A recent study based on the mathematical modeling conducted by Indian Council of Medical Research (ICMR) has suggested the proper quarantine and isolation as the preventive measures for stopping the outbreak of COVID-19 through community transmission [11]. It appears that in view of this, country- wide lock-down has been declared in India since the midnight of the 23 rd March, 2020 in continuation to pre-lock down action of evacuating the possible places expected for people's gathering as well as the social awareness drive already started from the very beginning of February, 2020 which we welcome.

Appreciable contribution in the development of diagnostics, therapeutics and vaccines for this novel corona virus has been indicated [11] and based on some clinical investigations, an anti-malarial drug namely chloroquine phosphate has been reported to be having a certain curative effect on the COVID-19 [12]. ICMR has also recommended an anti-malarial drug namely Hydroxy-chloroquine for those individuals which are asymptomatic healthcare workers involved in the care of suspected or confirmed cases of COVID-19 and asymptomatic household contacts of laboratory confirmed cases [13]. The possibility of high risk-factor associated with these suggested drugs has not been ruled out which projected it to be a trial measure. No specific therapy and medicine for the treatment of COVID-19 has ever been reported to the best of our knowledge which inspired us to think on the natural products relying on the concept of particle-antiparticle theory of Nature implying that if there is a novel corona virus then there must be its anti-virus material in the Nature itself and we have obtained motivating results in the present research carried out with the help of available computational facility at our home during lock-down period.

We have selected total eleven bioactive natural compounds embedded spontaneously in *Anthocephalus Cadamba* which has been reported to be a miraculous tree having crucial significance in Hindu Mythology containing the largest number of phytochemicals and secondary metabolites having pharmacological and biological properties, however, the solubility and solvent-effect to be the point of concern [14]. The four bio molecules contained in the leaves of the said tree namely 7-hydroxy-6-methoxy coumarian ($C_{10}H_8O_4$), Methyl ester of chlorogenic acid ($C_{17}H_{20}O_9$), Pentyle Ester of Chlorogenic Acid ($C_{21}H_{28}O_9$), D-Myo-Inositol ($C_7H_{14}O_6$); 07 biomolecules contained in fruits of the said tree namely Oleanic Acid ($C_{30}H_{48}O_3$), Ursolic Acid ($C_{30}H_{48}O_3$), Vallesiachotamine ($C_{21}H_{22}N_2O_3$), Iso-Vallesiachotamine ($C_{21}H_{22}N_2O_3$), Cadambine ($C_{27}H_{32}N_2O_{10}$), Vincosamide-N-Oxide ($C_{26}H_{31}N_2O_9$), Isodihydroamino-cadambine ($C_{26}H_{33}N_3O_7$) already reported to be isolated and characterized in Central Drug Research Institute, Lucknow, India [15-16] and the electronic properties of the few of them have already been studied by us computationally [17-21]; were screened as ligands to interact with the targeted SARS-COV-2 protease. The first two of the aforsaid compounds responded negatively but rest nine compounds exhibited the positive results which have been presented in this paper. We expect that the present study will offer a new dimension in developing the drug/vaccine for the COVID-19.

**Methodology**

The *In silico* optimized ligand-structure of the aforementioned eleven bioactive natural compounds were obtained as per reported approach of density functional theory at B3LYP/631+G (d, p) level [22-23] implemented through Gaussian 09 program-package [24]. We have used main protease of SARS-COV-2 retrieved from the RCSB protein data bank having PDB ID: 6Y84 [25] and PDB ID: 6LU7 [26] as potential target protein for the binding of these ligands obtained using molecular docking approach implemented through Autodock 4.2 program package [27-29]. The binding of these natural bioactive ligands with the PDB ID: 6CRV which is SARS Spike Glycoprotein for corona virus SARS-COV emerged in 2002 as a highly transmissible [30] has also been obtained with the help of same docking approach in order to investigate the consistency of the performance of these compounds with the class of viral protein. The binding of these compounds with the PDB ID: 6VXX which is the SARS-COV-2 spike glycoprotein for COVID-19 [31] has also been examined to illustrate the potential of these compounds to deactivate this novel corona virus at the receptor end.

**Results and Discussion**

The *In silico* optimized molecular structure of 09 bioactive natural compounds modeled using DFT-B3LY/6-31+G (d,p) level of theory which exhibited the property of inhibitor against main protease of COVID-19, have been displayed in figure 1. The binding affinity for all displayed molecules as ligand with the target protease of the SARS-Cov-2 virus (PDB ID: 6Y84 and 6LU7) as well as with the SARS Spike Glycoprotein for corona virus (PDB ID: 6CRV) have been depicted in table 1 to table 9. We observe a complete cycle for the binding of all these compounds with the said protease in maximum ten cluster run and the final free energy of binding is significant for all the ligand-protein interactions. The insignificant result for the binding of molecule no.8 (pentyl ester of chlorogenic acid) with 6LU7 and 6CRV receptors have been obtained where the same molecule exhibits a significant final free energy of binding with 6Y84 protease of COVID-19 as obvious from the table 8. The significant binding of these natural compounds with the said protease results in the conformational changes in the main protease of SARS-COV-2 as displayed in figure 2 to figure 10. It is, therefore, predicted that the naturally occurring bio molecules displayed in figure 1 may be explored as potential inhibitors for the COVID-19 protease. The docking results of these molecules are presented in decreasing order of the binding affinity i.e. the binding affinity with the target protein is largest for molecule 1 followe by molecule 2,3,4,5,6,7,8 and 9.

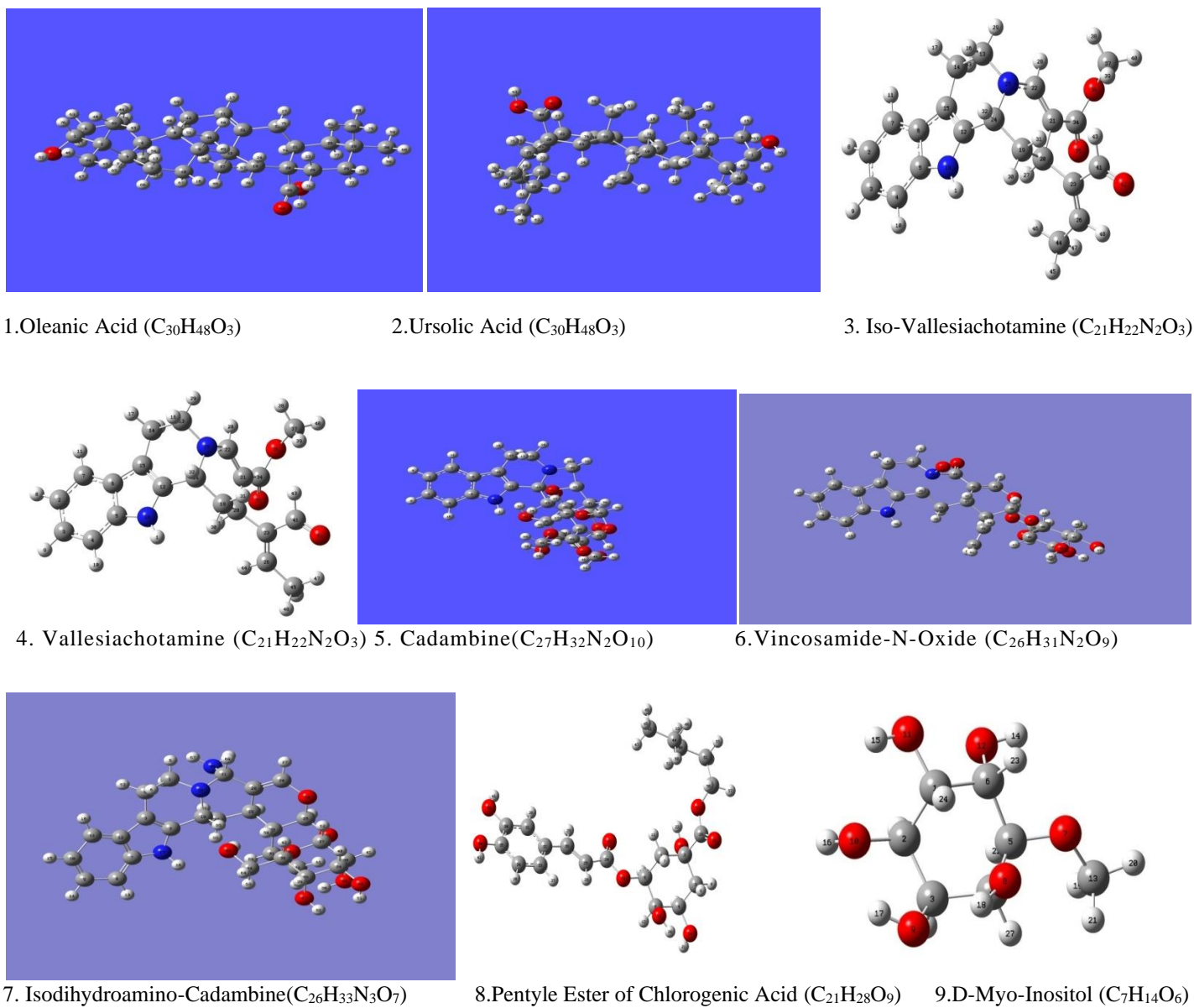

1. Oleanic Acid ($C_{30}H_{48}O_3$)   2. Ursolic Acid ($C_{30}H_{48}O_3$)   3. Iso-Vallesiachotamine ($C_{21}H_{22}N_2O_3$)

4. Vallesiachotamine ($C_{21}H_{22}N_2O_3$)  5. Cadambine($C_{27}H_{32}N_2O_{10}$)  6. Vincosamide-N-Oxide ($C_{26}H_{31}N_2O_9$)

7. Isodihydroamino-Cadambine($C_{26}H_{33}N_3O_7$)   8. Pentyle Ester of Chlorogenic Acid ($C_{21}H_{28}O_9$)   9. D-Myo-Inositol ($C_7H_{14}O_6$)

Figure 1: The optimized molecular structure of Nine Bioactive Natural Compounds *reported to be found in fruits (*molecule no 1 to 7*) and leaves (*molecule no 8 to 9*) of *Anthocephalus cadamba* [* 14-16]. Blue balls: N-atoms, black balls: C-atoms, white balls: H-atoms

The significantly negative value of the final free energy of binding obtained through the molecular docking of these compounds with the SARS-COV-2 protease reveals that these molecules may be explored as potential inhibitor against COVID-19. Since these compounds are naturally occurring, hence they do not cost for their synthesis and bears less or nil side-effect which enables them to be explored as user-friendly drug. The solubility and the effect of the solvent which are usually associated with the phytochrmicals may be the point of attention while using these compounds as drug candidate. The other chemical structure may also be derived from these bioactive natural compounds to further design potential drug agent for COVID-19.

Table 1: Binding affinity of Oleanic Acid ($C_{30}H_{48}O_3$) with the target proteins

| Cluster Rank | SARS-COV-2 protease (PDB ID: 6Y84) | | SARS-COV-2 protease (PDB ID: 6LU7) | | SARS-COV Spike Glycoprotein (PDB ID: 6CRV) | |
|---|---|---|---|---|---|---|
| | Free Energy of Binding | Inhibition Constant | Free Energy of Binding | Inhibition Constant | Free Energy of Binding | Inhibition Constant |
| 1 | -11.28 kcal/mol | 5.38nM | -6.11 kcal/mol | 33.45μM | -5.90 kcal/mol | 47.30μM |
| 2 | -9.78 kcal/mol | 67.90nM | -5.83 kcal/mol | 52.86μM | -5.73 kcal/mol | 63.60μM |
| 3 | -9.60 kcal/mol | 91.21nM | -5.70 kcal/mol | 66.07μM | -5.41 kcal/mol | 107.78μM |
| 4 | -9.10 kcal/mol | 213.24nM | -5.59 kcal/mol | 79.53μM | -4.91 kcal/mol | 249.99μM |
| 5 | --------- | --------- | -5.57 kcal/mol | 82.47μM | -4.39 kcal/mol | 607.04μM |
| 6 | --------- | --------- | -5.44 kcal/mol | 102.54μM | -4.28 kcal/mol | 723.95μM |
| 7 | --------- | --------- | -5.08 kcal/mol | 189.25μM | -4.25 kcal/mol | 768.82μM |
| 8 | --------- | --------- | -4.89 kcal/mol | 258.48μM | --------- | --------- |
| 9 | --------- | --------- | -4.77 kcal/mol | 318.64μM | --------- | --------- |
| 10 | --------- | --------- | --------- | --------- | --------- | --------- |

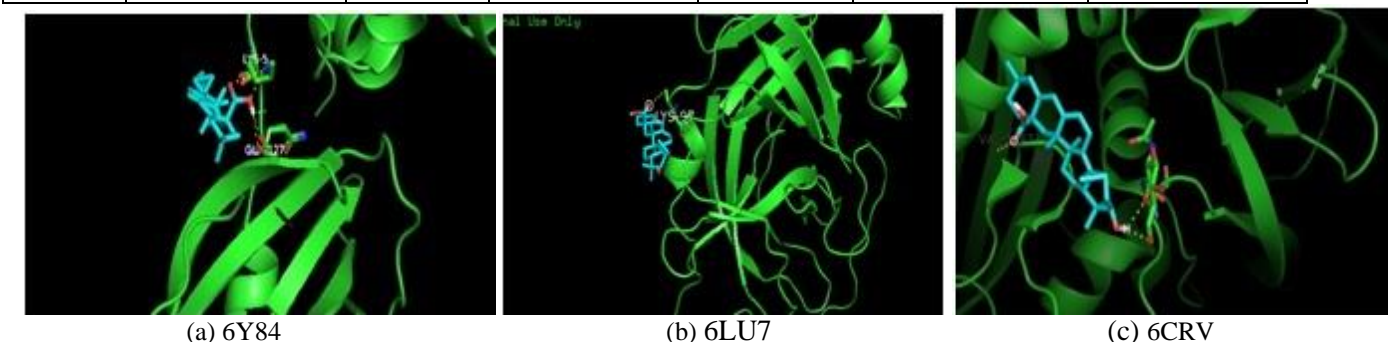

(a) 6Y84　　　　　　　　　　　(b) 6LU7　　　　　　　　　　　(c) 6CRV

Figure 2: Conformational changes observed due the binding of ligand-1 (Oleanic Acid) with the COVID-19 protease and S-protein receptor

Table 2: Binding affinity of Ursolic Acid ($C_{30}H_{48}O_3$) with the target proteins

| Cluster Rank | SARS-COV-2 protease (PDB ID: 6Y84) | | SARS-COV-2 protease (PDB ID: 6LU7) | | SARS-COV Spike Glycoprotein (PDB ID: 6CRV) | |
|---|---|---|---|---|---|---|
| | Free Energy of Binding | Inhibition Constant | Free Energy of Binding | Inhibition Constant | Free Energy of Binding | Inhibition Constant |
| 1 | -10.94 kcal/mol | 9.63nM | -6.40 kcal/mol | 20.38μM | -6.96 kcal/mol | 7.88μM |
| 2 | -10.30 kcal/mol | 28.35nM | -6.34 kcal/mol | 22.70μM | -6.81 kcal/mol | 10.26μM |
| 3 | -9.74 kcal/mol | 72.73nM | -6.00 kcal/mol | 40.32μM | -6.76 kcal/mol | 11.17μM |
| 4 | -9.72 kcal/mol | 74.54nM | -5.89 kcal/mol | 47.97μM | -6.61 kcal/mol | 14.24μM |
| 5 | -9.36 kcal/mol | 137.19nM | -5.34 kcal/mol | 122.36μM | -6.30 kcal/mol | 24.24μM |
| 6 | -8.90 kcal/mol | 299.83nM | -4.96 kcal/mol | 232.96μM | -6.19 kcal/mol | 29.17μM |
| 7 | --------- | --------- | --------- | --------- | -6.14 kcal/mol | 31.59μM |
| 8 | --------- | --------- | --------- | --------- | -6.08 kcal/mol | 34.90μM |
| 9 | --------- | --------- | --------- | --------- | -5.91 kcal/mol | 46.80μM |
| 10 | --------- | --------- | --------- | --------- | -5.86 kcal/mol | 50.93μM |

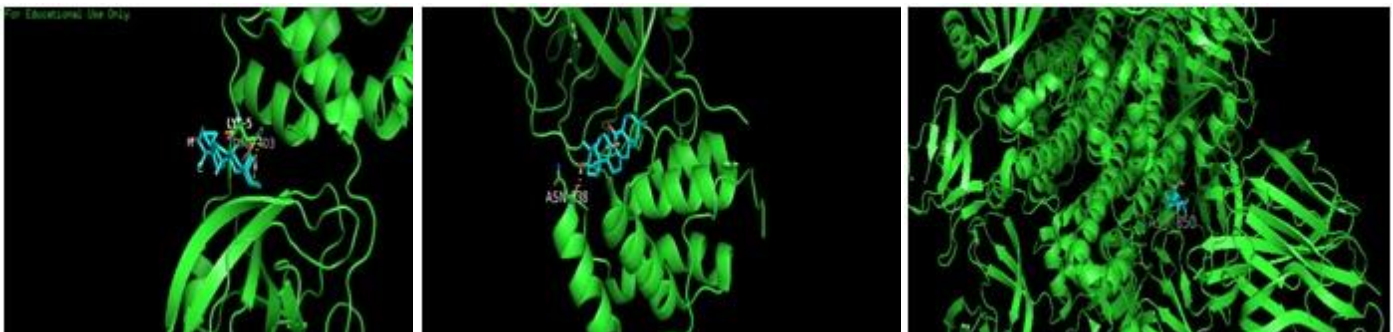

(a) 6Y84     (b) 6LU7     (c) 6CRV

Figure 3: Conformational changes observed due the binding of ligand-2 (Ursolic Acid) with the COVID-19 protease and S-protein receptor

Table 3: Binding affinity of Iso-Vallesiachotamine ($C_{21}H_{22}N_2O_3$) with the target proteins

| Cluster Rank | SARS-COV-2 protease (PDB ID: 6Y84) | | SARS-COV-2 protease (PDB ID: 6LU7) | | SARS-COV Spike Glycoprotein (PDB ID: 6CRV) | |
|---|---|---|---|---|---|---|
| | Free Energy of Binding | Inhibition Constant | Free Energy of Binding | Inhibition Constant | Free Energy of Binding | Inhibition Constant |
| 1 | -9.55 kcal/mol | 99.42nM | -6.00 kcal/mol | 39.82μM | -5.16 kcal/mol | 165.89μM |
| 2 | -7.93 kcal/mol | 1.53μM | -5.71 kcal/mol | 65.58μM | -5.00 kcal/mol | 214.52μM |
| 3 | -7.63 kcal/mol | 2.54μM | -5.70 kcal/mol | 66.53μM | -4.94 kcal/mol | 237.57μM |
| 4 | -7.61 kcal/mol | 2.63μM | -5.35 kcal/mol | 20.56μM | -4.81 kcal/mol | 296.22μM |
| 5 | -7.45 kcal/mol | 3.43μM | -5.11 kcal/mol | 178.30μM | -4.75 kcal/mol | 332.16μM |
| 6 | -6.51 kcal/mol | 16.94μM | -4.86 kcal/mol | 275.30μM | -4.70 kcal/mol | 357.55μM |
| 7 | --------- | --------- | --------- | --------- | -4.68 kcal/mol | 370.48μM |
| 8 | --------- | --------- | --------- | --------- | -4.45 kcal/mol | 548.28μM |
| 9 | --------- | --------- | --------- | --------- | -4.36 kcal/mol | 640.89μM |
| 10 | --------- | --------- | --------- | --------- | -4.03 kcal/mol | 1.11mM |

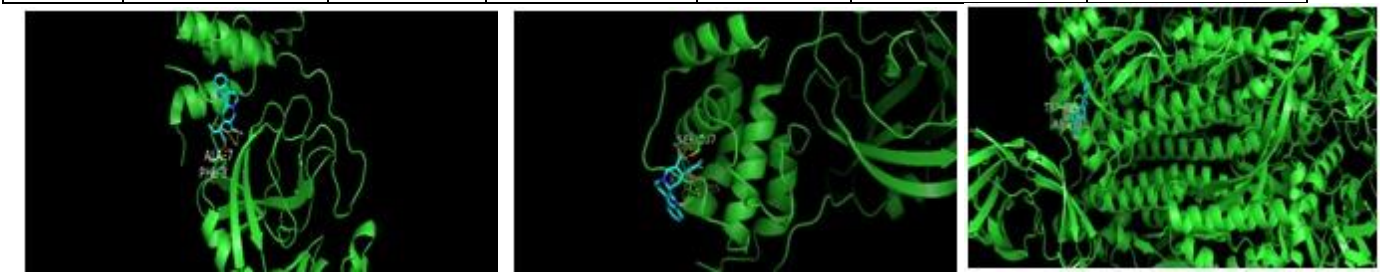

(a) 6Y84     (b) 6LU7     (c) 6CRV

Figure 4: Conformational changes observed due the binding of ligand-3 (Iso-Vallesiachotamine) with the COVID-19 protease and S-protein receptor

Table 4: Binding affinity of Vallesiachotamine($C_{21}H_{22}N_2O_3$) with the target proteins

| Cluster Rank | SARS-COV-2 protease (PDB ID: 6Y84) | | SARS-COV-2 protease (PDB ID: 6LU7) | | SARS-COV Spike Glycoprotein (PDB ID: 6CRV) | |
|---|---|---|---|---|---|---|
| | Free Energy of Binding | Inhibition Constant | Free Energy of Binding | Inhibition Constant | Free Energy of Binding | Inhibition Constant |
| 1 | -9.50 kcal/mol | 109.66nM | -4.98 kcal/mol | 224.33μM | -6.35 kcal/mol | 22.14μM |
| 2 | -9.32 kcal/mol | 148.51nM | -4.80 kcal/mol | 302.86μM | -6.04 kcal/mol | 37.63μM |
| 3 | -8.77 kcal/mol | 371.00nM | -4.43 kcal/mol | 564.68μM | -5.52 kcal/mol | 89.42μM |
| 4 | -8.33 kcal/mol | 784.52nM | -4.31 kcal/mol | 693.59μM | -5.05 kcal/mol | 197.72μM |
| 5 | -6.91 kcal/mol | 8.64μM | -4.30 kcal/mol | 699.66μM | -4.88 kcal/mol | 262.69μM |
| 6 | -6.87 kcal/mol | 9.28μM | -4.28 kcal/mol | 723.92μM | -4.77 kcal/mol | 320.94μM |
| 7 | -6.62 kcal/mol | 13.94μM | -3.94 kcal/mol | 1.30mM | -4.36 kcal/mol | 633.00μM |
| 8 | --------- | --------- | -3.73 kcal/mol | 1.84mM | -4.25 kcal/mol | 769.77μM |
| 9 | --------- | --------- | -2.84 kcal/mol | 8.22mM | -4.02 kcal/mol | 1.12mM |
| 10 | --------- | --------- | --------- | --------- | -3.90 kcal/mol | 1.39mM |

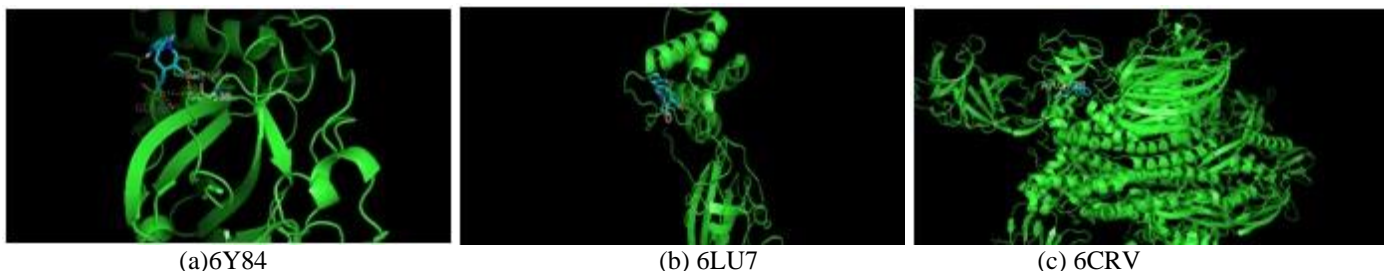

(a) 6Y84      (b) 6LU7      (c) 6CRV

Figure 5: Conformational changes observed due the binding of ligand-4 (Vallesiachotamine) with the COVID-19 protease and S-protein receptor

Table 5: Binding affinity of Cadambine($C_{27}H_{32}N_2O_{10}$) with the target proteins

| Cluster Rank | SARS-COV-2 protease (PDB ID: 6Y84) | | SARS-COV-2 protease (PDB ID: 6LU7) | | SARS-COV Spike Glycoprotein (PDB ID: 6CRV) | |
|---|---|---|---|---|---|---|
| | Free Energy of Binding | Inhibition Constant | Free Energy of Binding | Inhibition Constant | Free Energy of Binding | Inhibition Constant |
| 1 | -8.85 kcal/mol | 323.97nM | -6.93 kcal/mol | 13.86μM | -5.35 kcal/mol | 118.85μM |
| 2 | -7.57 kcal/mol | 2.81μM | -4.80 kcal/mol | 301.24μM | -4.83 kcal/mol | 287.35μM |
| 3 | -7.46 kcal/mol | 3.39μM | -4.44 kcal/mol | 553.34μM | -4.65 kcal/mol | 391.37μM |
| 4 | -7.18 kcal/mol | 5.45μM | -4.26 kcal/mol | 756.51μM | -4.56 kcal/mol | 451.55μM |
| 5 | -6.79 kcal/mol | 10.57μM | -4.24 kcal/mol | 779.20μM | -4.33 kcal/mol | 673.01μM |
| 6 | -6.22 kcal/mol | 27.54μM | -4.05 kcal/mol | 1.07mM | -4.24 kcal/mol | 783.98μM |
| 7 | -6.15 kcal/mol | 30.85μM | -4.03 kcal/mol | 1.11mM | -3.78 kcal/mol | 1.71mM |
| 8 | -5.85 kcal/mol | 51.89μM | -3.72 kcal/mol | 1.88mM | -3.50 kcal/mol | 2.70mM |
| 9 | -5.72 kcal/mol | 64.44μM | -3.33 kcal/mol | 3.60mM | -3.06 kcal/mol | 5.75mM |
| 10 | -5.45 kcal/mol | 100.34μM | -3.31 kcal/mol | 3.77mM | -2.88 kcal/mol | 7.78mM |

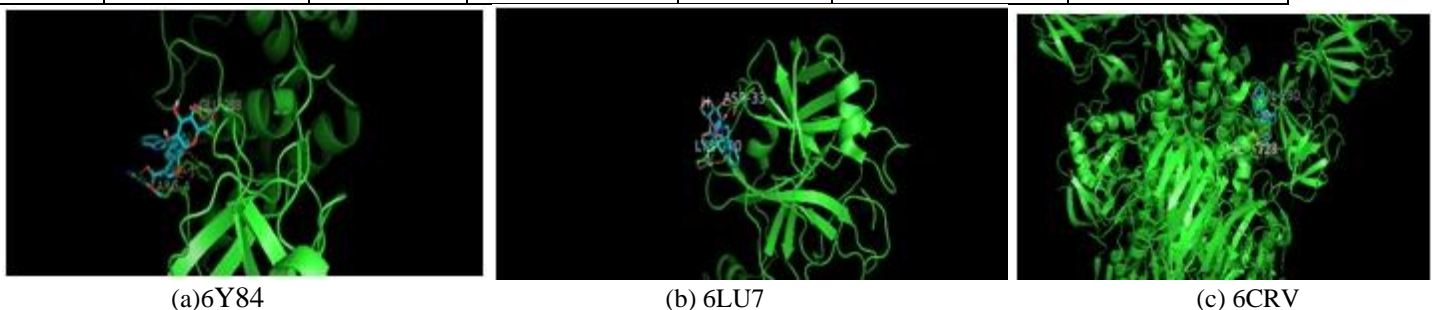

(a) 6Y84      (b) 6LU7      (c) 6CRV

Figure 6: Conformational changes observed due the binding of ligand-5 (Cadambine) with the COVID-19 protease and S-protein receptor

Table 6: Binding affinity of Vincosamide-N-Oxide ($C_{26}H_{31}N_2O_9$) with the target proteins

| Cluster Rank | SARS-COV-2 protease (PDB ID: 6Y84) | | SARS-COV-2 protease (PDB ID: 6LU7) | | SARS-COV Spike Glycoprotein (PDB ID: 6CRV) | |
|---|---|---|---|---|---|---|
| | Free Energy of Binding | Inhibition Constant | Free Energy of Binding | Inhibition Constant | Free Energy of Binding | Inhibition Constant |
| 1 | -8.81 kcal/mol | 348.65nM | -5.81 kcal/mol | 55.52μM | -4.59 kcal/mol | 434.05μM |
| 2 | -8.77 kcal/mol | 370.30nM | -4.78 kcal/mol | 312.98μM | -4.47 kcal/mol | 532.26μM |
| 3 | -8.41 kcal/mol | 686.48nM | -4.49 kcal/mol | 511.38μM | -4.28 kcal/mol | 730.11μM |

| 4 | -8.08 kcal/mol | 1.19μM | -3.93 kcal/mol | 1.31mM | -4.05 kcal/mol | 1.07mM |
| 5 | -7.92 kcal/mol | 1.58μM | -3.89 kcal/mol | 1.40mM | -4.00 kcal/mol | 1.17mM |
| 6 | -7.70 kcal/mol | 2.26μM | -3.75 kcal/mol | 1.79mM | -3.49 kcal/mol | 2.78M |
| 7 | -7.22 kcal/mol | 5.08μM | -3.48 kcal/mol | 2.83mM | -3.44 kcal/mol | 3.02mM |
| 8 | -6.26 kcal/mol | 25.76μM | -3.41 kcal/mol | 3.15mM | -3.38 kcal/mol | 3.31mM |
| 9 | -5.08 kcal/mol | 187.55μM | -3.33 kcal/mol | 3.64mM | -3.29 kcal/mol | 3.91mM |
| 10 | --------- | --------- | -3.03 kcal/mol | 5.96mM | -3.04 kcal/mol | 5.93mM |

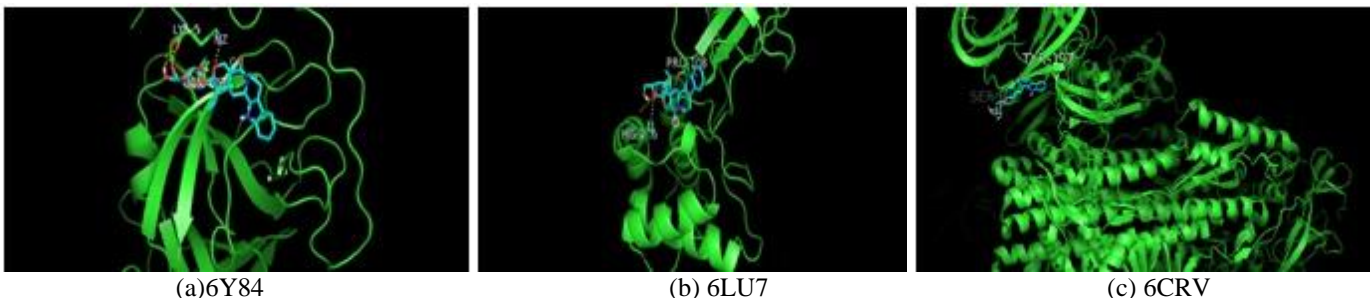

(a) 6Y84      (b) 6LU7      (c) 6CRV

Figure 7: Conformational changes observed due the binding of ligand-6 (Vincosamide-N-Oxide) with the COVID-19 protease and S-protein receptor

Table 7: Binding affinity of Isodihydroamino-cadambine($C_{26}H_{33}N_3O_7$) with the target proteins

| Cluster Rank | SARS-COV-2 protease (PDB ID: 6Y84) | | SARS-COV-2 protease (PDB ID: 6LU7) | | SARS-COV Spike Glycoprotein (PDB ID: 6CRV) | |
|---|---|---|---|---|---|---|
| | Free Energy of Binding | Inhibition Constant | Free Energy of Binding | Inhibition Constant | Free Energy of Binding | Inhibition Constant |
| 1 | -8.57 kcal/mol | 520.33nM | -4.85 kcal/mol | 277.88μM | -4.10 kcal/mol | 922.19μM |
| 2 | -7.88 kcal/mol | 1.67μM | -4.10 kcal/mol | 990.69μM | -3.90 kcal/mol | 1.39mM |
| 3 | -6.49 kcal/mol | 17.43μM | -4.08 kcal/mol | 1.02mM | -3.53 kcal/mol | 2.60mM |
| 4 | -5.27 kcal/mol | 137.23μM | -3.09 kcal/mol | 5.42mM | -3.44 kcal/mol | 2.99mM |
| 5 | -5.11 kcal/mol | 181.03μM | -2.92 kcal/mol | 7.26mM | -3.31 kcal/mol | 3.73mM |
| 6 | -4.76 kcal/mol | 324.79μM | -2.58 kcal/mol | 12.88mM | -3.00 kcal/mol | 6.28mM |
| 7 | -4.21 kcal/mol | 816.29μM | -2.53 kcal/mol | 14.04mM | -2.97 kcal/mol | 6.63mM |
| 8 | -3.90 kcal/mol | 1.39mM | -2.51 kcal/mol | 14.54mM | -2.88 kcal/mol | 7.71mM |
| 9 | --------- | --------- | -1.04 kcal/mol | 172.81mM | -2.84 kcal/mol | 8.27mM |
| 10 | --------- | --------- | --------- | --------- | -2.69 kcal/mol | 10.75mM |

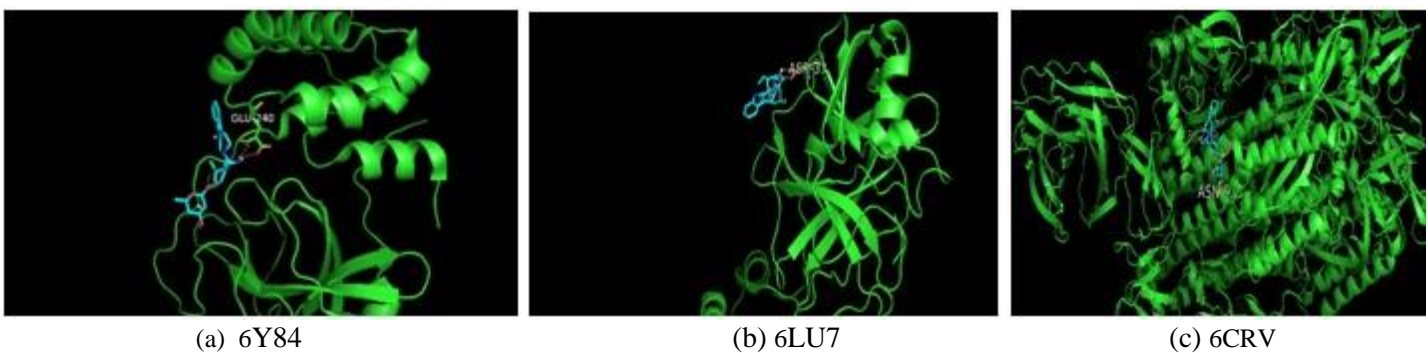

(a) 6Y84      (b) 6LU7      (c) 6CRV

Figure 8: Conformational changes observed due the binding of ligand-7 (Isodihydroamino-cadambine) with the COVID-19 protease and S-protein receptor

Table 8: Binding affinity of Pentyle ester of chlorogenic acid ($C_{21}H_{28}O_9$) with the target proteins

| Cluster Rank | SARS-COV-2 protease (PDB ID: 6Y84) | | SARS-COV-2 protease (PDB ID: 6LU7) | | SARS-COV Spike Glycoprotein (PDB ID: 6CRV) | |
|---|---|---|---|---|---|---|
| | Free Energy of Binding | Inhibition Constant | Free Energy of Binding | Inhibition Constant | Free Energy of Binding | Inhibition Constant |
| 1 | -5.61 kcal/mol | 77.71μM | -1.95 kcal/mol | 37.57mM | -2.35 kcal/mol | 19.10mM |
| 2 | -5.10 kcal/mol | 182.54μM | -1.52 kcal/mol | 76.24mM | -1.77 kcal/mol | 50.49mM |
| 3 | -5.06 kcal/mol | 194.58μM | -1.46 kcal/mol | 85.21mM | -1.66 kcal/mol | 60.43mM |
| 4 | -5.05 kcal/mol | 198.85μM | -1.03 kcal/mol | 176.88mM | -1.64 kcal/mol | 62.51mM |
| 5 | -4.18 kcal/mol | 869.07μM | -0.86 kcal/mol | 233.66mM | -1.37 kcal/mol | 99.72mM |
| 6 | -3.40 kcal/mol | 3.21mM | -0.61 kcal/mol | 356.69mM | -1.23 kcal/mol | 126.32mM |
| 7 | -3.30 kcal/mol | 3.80mM | -0.42 kcal/mol | 490.95mM | -1.03 kcal/mol | 176.85mM |
| 8 | -3.15 kcal/mol | 4.87mM | -0.37 kcal/mol | 531.76mM | -0.20 kcal/mol | 713.11mM |
| 9 | -3.03 kcal/mol | 6.04mM | -0.06 kcal/mol | 903.01mM | -0.10 kcal/mol | 840.36mM |
| 10 | -2.86 kcal/mol | 8.01mM | +0.05 kcal/mol | ------------ | -0.00 kcal/mol | 991.66mM |

(a) 6Y84      (b) 6LU7      (c) 6CRV

Figure 9: Conformational changes observed due the binding of ligand-8 (Pentyle Ester of Chlorogenic Acid) with the COVID-19 protease and S-protein receptor

Table 9: Binding affinity of D-Myo-Inositol ($C_7H_{14}O_6$) with the target proteins

| Cluster Rank | SARS-COV-2 protease (PDB ID: 6Y84) | | SARS-COV-2 protease (PDB ID: 6LU7) | | SARS-COV Spike Glycoprotein (PDB ID: 6CRV) | |
|---|---|---|---|---|---|---|
| | Free Energy of Binding | Inhibition Constant | Free Energy of Binding | Inhibition Constant | Free Energy of Binding | Inhibition Constant |
| 1 | -5.51 kcal/mol | 90.67μM | -2.22 kcal/mol | 23.63mM | -3.11 kcal/mol | 5.22mM |
| 2 | -5.06 kcal/mol | 195.69μM | -2.05 kcal/mol | 31.17mM | -2.77 kcal/mol | 9.34mM |
| 3 | -4.30 kcal/mol | 702.05μM | -1.86 kcal/mol | 43.37mM | -2.64 kcal/mol | 11.57mM |
| 4 | -3.56 kcal/mol | 2.47μM | -1.74 kcal/mol | 53.02mM | -2.53 kcal/mol | 13.88mM |
| 5 | -3.04 kcal/mol | 5.86mM | -1.72 kcal/mol | 54.66mM | -2.32 kcal/mol | 20.09mM |
| 6 | -3.00 kcal/mol | 6.30mM | -1.53 kcal/mol | 75.47mM | -2.30 kcal/mol | 20.46mM |
| 7 | -2.93 kcal/mol | 7.09mM | -1.38 kcal/mol | 98.07mM | -2.11 kcal/mol | 28.50mM |
| 8 | --------- | --------- | -1.10 kcal/mol | 154.91mM | -1.98 kcal/mol | 35.50mM |
| 9 | --------- | --------- | -1.07 kcal/mol | 164.03mM | -1.94 kcal/mol | 37.54mM |
| 10 | --------- | --------- | --------- | --------- | -1.56 kcal/mol | 71.41mM |

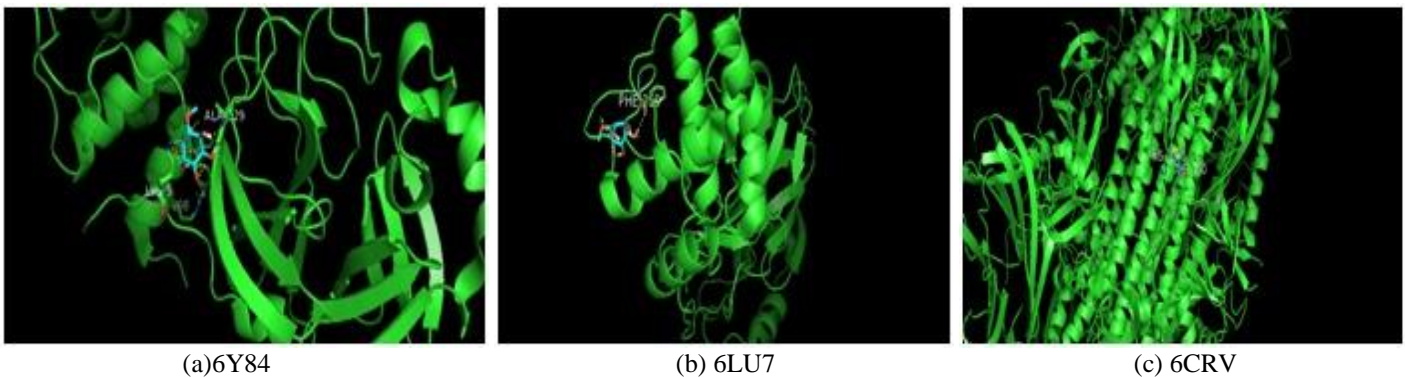

(a) 6Y84  (b) 6LU7  (c) 6CRV

Figure 10: Conformational changes observed due the binding of ligand-9 (D-Myo-Inositol) with the COVID-19 protease and S-protein receptor

The significantly negative value of free energy of binding of these molecules with PDB ID: 6VXX (SARS-COV-2 spike glycoprotein for COVID-19) in ligand-protein interaction as depicted in table 10 and 11 reveals that these molecules or their structural-derivatives may stop the replication of the virus at the receiver end in the human body. The results regarding binding affinity of the some of these compounds and corresponding conformational changes occurred in the target protein are presented in figure 11and 12.

Table 10: Binding affinity in SARS-COV-2 spike glycoprotein interaction

| | Oleanic Acid ($C_{30}H_{48}O_3$)-6VXX interaction | | Ursolic Acid ($C_{30}H_{48}O_3$) -6VXX interaction | |
|---|---|---|---|---|
| Cluster Rank | Free Energy of Binding | Inhibition Constant | Free Energy of Binding | Inhibition Constant |
| 1 | -6.76 kcal/mol | 11.05μM | -7.15 kcal/mol | 5.73μM |
| 2 | -6.54 kcal/mol | 16.17μM | -7.07 kcal/mol | 6.53μM |
| 3 | -6.41 kcal/mol | 19.85μM | -6.74 kcal/mol | 11.56μM |
| 4 | -6.41 kcal/mol | 19.90μM | -6.72 kcal/mol | 11.78μM |
| 5 | -6.12 kcal/mol | 32.71μM | -6.48 kcal/mol | 17.64μM |
| 6 | -6.04 kcal/mol | 37.45μM | -6.39 kcal/mol | 20.63μM |
| 7 | -5.80 kcal/mol | 56.45μM | -6.29 kcal/mol | 24.67μM |
| 8 | -5.70 kcal/mol | 66.58μM | -6.22 kcal/mol | 27.76μM |
| 9 | -5.45 kcal/mol | 101.12μM | -6.20 kcal/mol | 28.40 μM |
| 10 | | | -5.81 kcal/mol | 55.41μM |

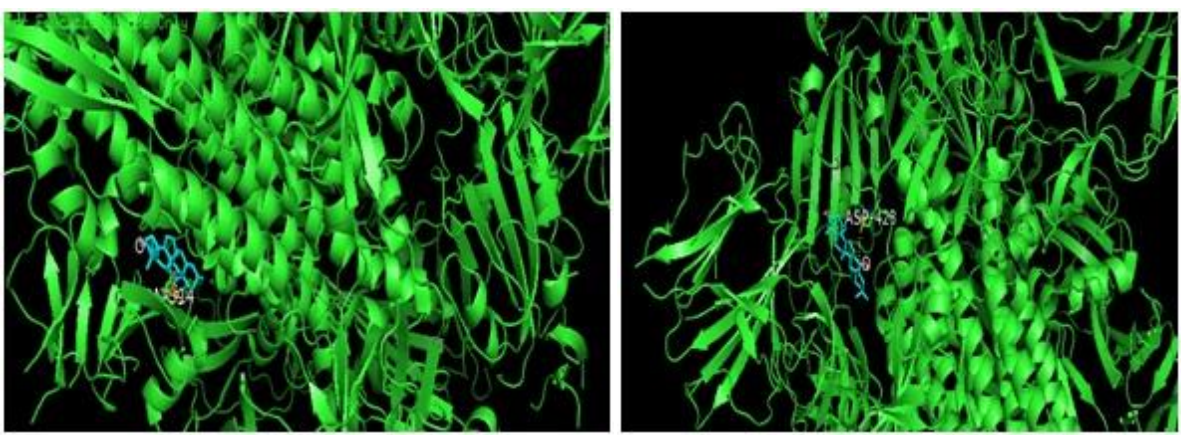

Figure 11 (a) Ligand [Oleanic Acid] - Protein [6VXX] interaction  (b) Ligand [Ursolic Acid ] - Protein [6VXX] interaction

Table 11: Binding affinity in SARS-COV-2 spike glycoprotein interaction

| | Iso-Vallesiachotamine (C$_{21}$H$_{22}$N$_2$O$_3$)-6VXX interaction | | Isodihydroamino-cadambine (C$_{26}$H$_{33}$N$_3$O$_7$) -6VXX interaction | |
|---|---|---|---|---|
| **Cluster Rank** | **Free Energy of Binding** | **Inhibition Constant** | **Free Energy of Binding** | **Inhibition Constant** |
| 1 | -5.18 kcal/mol | 160.30μM | -5.36 kcal/mol | 118.39μM |
| 2 | -5.12 kcal/mol | 177.97μM | -3.88 kcal/mol | 1.44mM |
| 3 | -5.01 kcal/mol | 211.92μM | -3.56 kcal/mol | 2.46mM |
| 4 | -4.76 kcal/mol | 322.60μM | -3.48 kcal/mol | 2.80mM |
| 5 | -4.72 kcal/mol | 344.73μM | -3.35 kcal/mol | 3.49mM |
| 6 | -4.55 kcal/mol | 462.95μM | -3.22 kcal/mol | 4.39mM |
| 7 | -4.32 kcal/mol | 676.74μM | -2.88 kcal/mol | 7.78mM |
| 8 | -4.25 kcal/mol | 762.33μM | -2.65 kcal/mol | 11.39mM |
| 9 | -3.58 kcal/mol | 2.37mM | -2.62 kcal/mol | 12.08mM |
| 10 | -3.40 kcal/mol | 3.20mM | -2.53 kcal/mol | 14.00mM |

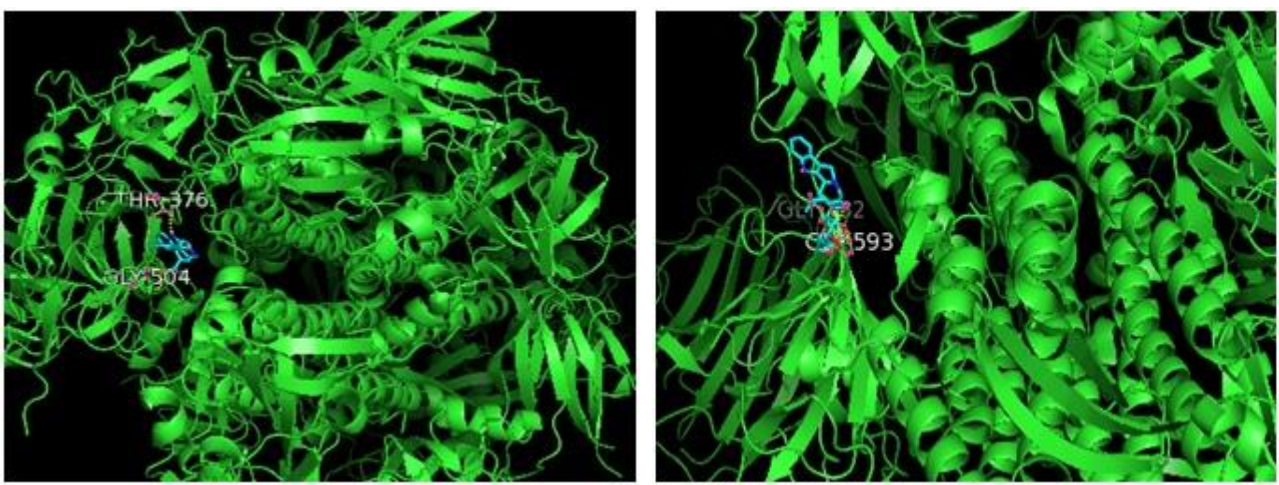

Figure 12 (a) Ligand [Iso-Vallesiachotamine] - Protein [6VXX] interaction  (b) Ligand [Isodihydroamino-cadambine] - Protein [6VXX] interaction

Similar types of significant binding affinity have been obtained in all the SARS-COV-2 spike glycoprotein – ligand (studied compounds) interactions whose results are available on demand but not depicted due to space limitations.

**Concluding Remarks**

We have carried out a schematic *In Silico* investigation applying density functional theory and molecular docking approach on the nine naturally occurring bioactive compounds listed in figure 1 spontaneously found in the leaves and fruits of *Anthocephalus cadamba*. On the basis of the binding affinity, we have found these compounds as potential inhibitors against the COVID-19 having final free energy of binding in the order of molecule 1>2>3>4>5>6>7>8>9. These molecules have also emerged as potential inhibitors against the SARS-COV-2 spike glycoprotein for COVID-19 in our investigation. These compounds may be explored as drug candidate and their molecular structure may be exploited to develop a vaccine for COVID-19. In-vivo study is invited on these compounds for developing user friendly drug for COVID-19. The further study on derived structures of these compounds is going on in our laboratory.

**Acknowledgements**

Authors are thankful to Professor Neeraj Misra, Department of Physics University of Lucknow, India for permitting us to use his computational facility. Thanks are due to Dr. D.P. Mishra, former Research fellow, Central Drug Research Institute, Lucknow, India for immense help regarding the information about bioactive natural products.